\documentclass[aps,prd,reprint,showpacs,longbibliography,superscriptaddress,groupedaddres,
titlepage,nofootinbib]{revtex4-1} 

\usepackage{amsthm}
\usepackage{amssymb}   
\usepackage{mathtools} 
\usepackage{hyperref}
\hypersetup{colorlinks=true,linkcolor=blue,urlcolor=blue,citecolor=blue}
\usepackage{accents}
\usepackage{tensor}
\usepackage[cal=boondox]{mathalfa}
\usepackage{lipsum}
\usepackage{relsize}
\usepackage{color}
\usepackage{comment}
\usepackage{nameref}
\usepackage[T1]{fontenc}
\newcommand{\ovg}[1]{\stackrel{(\gamma)}{#1}}

\begin{document}
	
	\title{$SU(1,1)$ Barbero-like variables derived from Holst action}

	\author{Merced Montesinos}
	\email{merced@fis.cinvestav.mx}
	\author{Jorge Romero}%
	\email{ljromero@fis.cinvestav.mx}
	\author{Ricardo Escobedo}
	\email{rescobedo@fis.cinvestav.mx}
	\affiliation{%
		Departamento de F\'{i}sica, Cinvestav, Avenida Instituto Polit\'{e}cnico Nacional 2508, \\
		San Pedro Zacatenco, 07360 Gustavo A. Madero, Ciudad de M\'exico, Mexico	
	}%
	
	\author{Mariano Celada}
	\email[]{mcelada@fis.cinvestav.mx}
	\affiliation{Departamento de F\'isica, Universidad Aut\'onoma Metropolitana Iztapalapa, San Rafael Atlixco 186, 09340 Ciudad de M\'exico, Mexico}
	
	\date{\today}
	
	\begin{abstract}
		We work on a spacetime manifold foliated by timelike leaves. In this setting, we explore the solution of the second-class constraints arising during the canonical analysis of the Holst action with a cosmological constant. The solution is given in a manifestly Lorentz-covariant fashion, and the resulting canonical formulation is expressed using several sets of real variables that are related to one another by canonical transformations. By applying a gauge fixing to this formulation, we obtain a description of gravity as an $SU(1,1)$ gauge theory that resembles the Ashtekar-Barbero formulation.
	\end{abstract}
	
	\maketitle
	
	\section{\label{sec:Intro} Introduction}
	
		The quest for a consistent quantum theory of gravity is perhaps the greatest endeavor of modern theoretical physics. One of the main approaches tackling this problem is loop quantum gravity~\cite{RovBook,ThieBook,Ashtekar0407,*Rovelli0812, *Rovelli1106}, which intends a canonical quantization of the gravitational field. This approach has provided not only deep insights into the quantum nature of gravity, but has also helped to establish some other alternative strategies for quantizing gravity in a nonperturbative framework such as the spinfoam models~\cite{perez2013} (see for instance Ref.~\cite{oriti_2009} for more proposals).
		
		The loop approach is based upon the Ashtekar-Barbero variables~\cite{Barbero9505}, which arise from the canonical analysis---in the time gauge---of the Holst action for general relativity~\cite{Holst}. This gauge condition plays an essential role in all the loop construction, since it breaks the noncompact Lorentz group down to its compact subgroup $SU(2)$ [or $SO(3)$], making things easier. The theory however includes a free parameter that plays no significant role in the space of solutions of the classical theory\footnote{It however manifests off shell in the classical theory. See Ref.~\cite{Montesinos1709}.} since it is introduced into the formalism as the coupling constant of a topological term~\cite{LiuPRD81}, but it unfortunately shows up in the spectra of quantum observables~\cite{Rovelli9505,*Ashtekar9700} and in the black hole entropy~\cite{Rovelli9610,*Ashtekar0000,*Meissner0410,*Agullo0805,*Engle1007}.~Thus, the meaning of the so-called Barbero-Immirzi or simply Immirzi parameter~\cite{Immirzicqg1410} is so far not clear, although some people think that its presence in the quantum theory might be a consequence of the use of the time gauge (others argue that the Immirzi parameter is a feature of quantum gravity itself). Because of this, there has been a great interest in dismissing the time gauge in favor of a description of the phase space of general relativity employing Lorentz-covariant variables~\cite{Barros0100,Alexcqg1720,*AlexVassprd644,*AlexLivprd674,NouiSIGMA72011,*Nouiprd84044002} (see also Ref.~\cite{Geiller2013,*Achour2014,*Achour2015} for a related discussion in a three-dimensional scenario).

		In view of this, in Ref.~\cite{Montesinos1801} the authors revisited the canonical analysis of the Holst action. Since second-class constraints arise during the process, they managed to solve them in a manifestly Lorentz-covariant fashion (In contrast, the approach of Ref.~\cite{Alexcqg1720}, although explicitly Lorentz covariant, employs the Dirac bracket.) The authors then obtained several sets of manifestly Lorentz-covariant variables for the phase space of general relativity that are related to one another by canonical transformations. In the time gauge, these variables easily and immediately lead to the Ashtekar-Barbero variables, and so the variables reported in Ref.~\cite{Montesinos1801} correspond to a Lorentz-covariant extension thereof. 
		
		Recently, in Ref.~\cite{Liu1706}, within the framework of Refs.~\cite{Barros0100,NouiSIGMA72011,*Nouiprd84044002} and in order to further explore the implications of the Immirzi parameter, a gauge fixing different from the time gauge was considered. In this gauge, which was termed the ``space gauge'', the four-dimensional Lorentz group is broken down to $SU(1,1)$ [or $SO(1,2)$, since they both share the same Lie algebra], obtaining a spacetime foliated by timelike three-manifolds. However, this approach is rather complicated, this in part due to the fact that the solution of the second-class constraints used in Ref.~\cite{Liu1706} is not adapted to the new gauge condition, as it is actually for the time gauge.

		In this work we explore, within the framework of Ref.~\cite{Montesinos1801}, the implications of the space gauge, and give a complete description of the resulting canonical theory. In particular, we highlight the simplicity of our approach as compared with that of Ref.~\cite{Liu1706}. As a result we obtain, by following two different paths, a set of real canonical variables and constraints resembling those of the Ashtekar-Barbero formalism, but this time with a remanent internal $SU(1,1)$ symmetry, as is the space gauge thought for.

		The outline of this paper is as follows. First, in Sec.~\ref{sec:HA} we sketch the canonical analysis of the Holst action when the ``evolution'' is considered along one of the spatial directions.~Then, in Sec.~\ref{sec:FCH} we solve the resulting second-class constraints in a manifestly Lorentz-covariant fashion and write down several sets of real phase-space variables related among them by canonical transformations. Later on, in Sec.~\ref{sec:GF} we exhibit, after the application of the space gauge, the resulting canonical variables and constraints. For the sake of completeness, in the Appendix we report the canonical formulation that emerges when the second-class constraints are solved in a nonexplicitly Lorentz-covariant fashion (but preserving full Lorentz invariance), and in Sec.~\ref{sec:AA} we adapt the space gauge to this framework and show that the canonical formulation of Sec.~\ref{sec:GF} also follows. At the end, we give some final remarks.

	\section{\label{sec:HA} Canonical Analysis}
	
		In this section we consider a spacetime foliation differing from the usual one. Instead of foliating spacetime with respect to a timelike direction, we do it with respect to a spacelike one. In consequence, the three-dimensional leaves that fill up spacetime are no longer spacelike but timelike (the case of a foliation by null hypersurfaces was addressed in Ref.~\cite{Alexandrov1503}). Thus, we assume that the spacetime manifold $M$ has the topology $M=\Sigma \times \mathbb{R}$, but this time $\Sigma$ is a noncompact timelike three-manifold; in analogy with the usual case, ``$\mathbb{R}$'' denotes here the spacelike direction along which $\Sigma$ ``evolves'', which means nothing but a mere change of timelike leaf. We denote this direction by $x^3$ (any of the other spatial directions can be equivalently chosen), and so the hypersurface $x^3=\text{constant}$ has the same topology of $\Sigma$ and its coordinates are designated by $x^a$, with $a=0,1,2$. The canonical analysis then proceeds as in the case of a timelike direction.
	
		We denote the internal indices by the capital letters of the middle of the alphabet $I, \, J, \,\ldots=\{i,3\}$, for $i=0,1,2$, which in turn are lowered or raised with Minkowski's metric $(\eta_{IJ})=\mbox{diag}(-1,1,1,1)$. We define the symmetrizer and the antisymmetrizer by $A_{(\mu\nu)} := (1/2) (A_{\mu\nu} + A_{\nu\mu})$ and $A_{[\mu\nu]} := (1/2) (A_{\mu\nu} - A_{\nu\mu})$, respectively. Furthermore, for any quantity taking values in the Lie algebra of the Lorentz group, $U_{IJ}=-U_{JI}$, we introduce its corresponding internal Hodge dual as $\star U_{IJ} = (1/2) \epsilon_{IJKL}U^{KL}$, with $\epsilon_{IJKL}$ being the totally antisymmetric Levi-Civita tensor that satisfies $\epsilon_{0123}=1$. Likewise, we also define a $\gamma$-valued quantity by
		\begin{equation}
			\ovg{U}_{IJ} := U_{IJ} + \dfrac{1}{\gamma}\star U_{IJ} = P_{IJ}{}^{KL} U_{KL}, 
		\end{equation}
		where $\gamma$ is the Immirzi parameter and $P_{IJ}{}^{KL} := \delta^{K}_{[I}\delta^{L}_{J]} + 
		(1/2\gamma) \epsilon_{IJ}{}^{KL}$. Finally, when working with tensor densities we will denote their negative (positive) weight with an equivalent amount of tildes ($\sim$) under (above) the corresponding tensor; beware that we will omit the tildes of tensors densities with weights lower than -1 or greater 2, but their weight will be specified somewhere else in the paper.
		
		The canonical analysis of the Holst action (with cosmological constant $\Lambda$) can then be performed in a Lorentz-covariant fashion along the same lines of Ref.~\cite{Barros0100} (see also Ref.~\cite{Peldan9400}). We end up (neglecting boundary terms) having a Hamiltonian action in the form
		\begin{equation}
				\label{Holst}
				S = \int_{\mathbb{R}} dx^3\int_{\Sigma}dV \left[\ovg{\tilde{\Pi}}{}^{aIJ} \partial_{3}{\omega_{aIJ}} - \tilde{H} \right],
		\end{equation}
		where $dV:=dx^0dx^1dx^2$, $(\omega_{aIJ},\,\ovg{\tilde{\Pi}}{}^{aIJ})$ [or equivalently $(\ovg{\omega}_{aIJ},\tilde{\Pi}^{aIJ})$] constitute the canonical variables and satisfy the relation $\{\omega_{aIJ}(x),\ovg{\tilde{\Pi}} {^{bKL}}(y)\}=\delta_{a}^{b}\delta_{[I}^{K}\delta_{J]}^{L}\delta^3(x,y)$, and $\tilde{H}$ is the Hamiltonian density given by
		\begin{equation}
				\tilde{H} :=  \lambda_{IJ} \tilde{\mathcal{G}}^{IJ} + N^{a}\tilde{\mathcal{V}}_{a} + \underaccent{\tilde}{N} \tilde{\tilde{\mathcal{H}}} + \underaccent{\tilde}{\phi}_{ab}\tilde{\tilde{\Phi}}^{ab} + \psi_{ab}\Psi^{ab},
		\end{equation}
		where $\lambda_{IJ}$, $N^{a}$, $\underaccent{\tilde}{N}$, $\underaccent{\tilde}{\phi}_{ab}$, and $\psi_{ab}$ (it has weight -2, so that $\Psi^{ab}$ has weight +3) are Lagrange multipliers that impose the following constraints\footnote{Although we should have rewritten the constraints entirely in terms of the canonical variables, either $(\omega_{aIJ},\ovg{\tilde{\Pi}}{}^{aIJ})$ or $(\ovg{\omega}_{aIJ},\tilde{\Pi}^{aIJ})$, the resulting expressions are rather clumsy, and so we decided to write them in their simplest form.}:
		\begin{subequations}
			\begin{eqnarray}
			&&\tilde{\mathcal{G}}^{IJ}:=D_{a}\!\!\ovg{\tilde{\Pi}}\!\!{}^{aIJ}\!\! = \partial_{a}\!{\ovg{\tilde{\Pi}\!\!}{}^{aIJ}}\! + 2 \omega_{a}{}^{[I|}{}_K \!\!\!\ovg{\tilde{\Pi}}{}^{aK|J]}\approx 0,\quad \label{Gauss}\\
			&&\tilde{\mathcal{V}}_a := \frac{1}{2}\ovg{\tilde{\Pi}}{}^{bIJ}F_{baIJ}\approx 0,\label{Vector} \\
			&&\tilde{\tilde{\mathcal{H}}}:=\frac{1}{2} \tilde{\Pi}^{aIK}\tilde{\Pi}^b{}_K{}^J\stackrel{(\gamma)}{F}\!\!{}{_{abIJ}} -\Lambda \tilde{\tilde{q}} \approx 0 ,\label{Scalar}\\
			&&\tilde{\tilde{\Phi}}^{ab} := \star\tilde{\Pi}^a{}_{IJ}\tilde{\Pi}^{bIJ} \approx 0, \label{Phi}\\
			&&\Psi^{ab} := \epsilon_{IJKL}\tilde{\Pi}^{(a|IM}\tilde{\Pi}^c{}_M{}^JD_c \tilde{\Pi}^{|b)KL}\approx 0\label{Psi},   	
			\end{eqnarray}
		\end{subequations}
		with $F_{ab IJ}:=2\left(\partial_{[a}\omega_{b]IJ}+\omega_{[a|IK}\omega_{|b]}{}^{K}{}_{J}\right)$ and $\tilde{\tilde{q}}:=\det(q_{ab})$, $q_{ab}$ being the induced metric on $\Sigma$, whose inverse $q^{ab}$ is related to $\tilde{\Pi}^{aIJ}$ by the relation 
		\begin{equation}
				\label{qPi}
				\tilde{\tilde{q}}q^{ab}= -\frac{1}{2}\tilde{\Pi}^{aIJ}\tilde{\Pi}^{b}{}_{IJ}.
		\end{equation}
		From the Poisson algebra among the constraints \eqref{Gauss}--\eqref{Psi}, we find that the Gauss $\tilde{\mathcal{G}}^{IJ}$, vector $\tilde{\mathcal{V}}_{a}$ [we may use instead the diffeomorphism constraint $\tilde{\mathcal{D}}_{a}:=\tilde{\mathcal{V}}_{a}+ (1/2) \omega_{aIJ}\tilde{\mathcal{G}}^{IJ}$], and scalar $\tilde{\tilde{\mathcal{H}}}$ constraints are first class (the Gauss constraint generates local Lorentz transformations, whereas the vector and scalar constraints generate spacetime diffeomorphisms), whereas $\tilde{\tilde{\Phi}}^{ab}$ and $\Psi^{ab}$ are second class. As a result, the theory propagates $(1/2)(2\times 18 - 2 \times 10 - 12)=2$ degrees of freedom, as expected for general relativity. The second-class constraints will be deal with in the next section.
		
		To close this section, it is worth pointing out that both the form of the action~\eqref{Holst} and the constraints \eqref{Gauss}-\eqref{Psi} are actually the same obtained when we perform the canonical analysis of the Holst action with respect to a timelike direction (that is, in the usual fashion). Thus, at this point in the Hamiltonian framework whether spacetime is foliated by timelike or spacelike hypersurfaces does not really matter. The direction ``$x^3$'' might play the role of a space direction as well as that of time. This is a manifestation of the fact that in a diffeomorphism-invariant theory there is no distinction between space and time. Nevertheless, we shall see that the introduction of a timelike or spacelike direction in the internal Minkowski space determines the nature of the metric induced on $\Sigma$, in turn giving a meaning to the spacetime direction ``$x^3$''.

	\section{\label{sec:FCH} First-class Hamiltonian}
		
		The above canonical theory is manifestly Lorentz covariant, but features the presence of second-class constraints. In order to obtain a Hamiltonian theory with first-class constraints solely while preserving local Lorentz invariance, in this section we explicitly solve the second-class constraints in such a way that this symmetry is neither split nor broken.

		
		\subsection{Solution to the second-class constraints}
		
		The constraint \eqref{Phi} consists of six restrictions for the 18 variables encoded in $\tilde{\Pi}^{aIJ}$, which means that its solution ought to be given in terms of 12 independent variables $\tilde{B}^{aI}$ (see also \cite{Montesinos1801,AshtLectures,Peldan9400})
			\begin{equation}
				\label{PiSol}
				\tilde{\Pi}^{a I J}=\epsilon\tilde{B}^{a [I}m^{J]},
			\end{equation}
		where $\epsilon = \pm 1$ due to the quadratic dependence on $\tilde{\Pi}^{a I J}$ of Eq.~\eqref{Phi}, and $m^I$ is an internal vector satisfying $m_{I}\tilde{B}^{aI}=0$ and $m_{I}m^{I}=\tau$, with $\tau=\pm 1$\footnote{Although the spacetime foliation was initially set up with respect to a spacelike direction, the Hamiltonian framework allows us to address timelike and spacelike directions at the same time.}. Since $m^I$ and $\tilde{B}^{aI}$ are orthogonal to each other, for $\tau=1$ ($\tau=-1$) or $m^I$ spacelike (timelike), $\tilde{B}^{aI}$ is timelike (spacelike). On the other hand, by defining $\tilde{\tilde{h}}{}^{ab}:=\tilde{B}^{a I}\tilde{B}^{b}{}_{I}$, the induced metric \eqref{qPi} takes the form $\tilde{\tilde{q}}q^{ab}=(-\tau/4)\tilde{\tilde{h}}^{ab}$, which implies $\tilde{\tilde{q}}{}^2=(-\tau/64)h$ for $h:=\det\tilde{\tilde{h}}{}^{ab}$ (of weight $+4$). Since the right-hand side of this equation must be positive, then the sign of $h$ must be the opposite of $\tau$, that is $h=-\tau|h|$. However, notice that the same relation does not fix the sign of $\tilde{\tilde{q}}$; we demand it to have the same sign as $h$, which implies $q^{ab}=2|h|^{-1/2}\tilde{\tilde{h}}{}^{ab}$. Therefore, for $\tau=1$ ($\tau=-1$) or $m^I$ spacelike (timelike), the induced metric on $\Sigma$ has Lorentzian (Euclidean) signature. Explicitly, $m_{I}$ has the form 
			\begin{equation} 
				\label{mI}
				m_{I}:=\frac{1}{6\sqrt{|h|}}\epsilon_{I J K L} \underaccent{\tilde}{\eta}_{a b c}\tilde{B}^{a J}\tilde{B}^{b K}\tilde{B}^{c L},
			\end{equation}
			
			\noindent
		where $\underaccent{\tilde}{\eta}_{a b c}$ is totally antisymmetric and satisfies $\underaccent{\tilde}{\eta}_{012}=1$. Denoting by $h_{ab}$ (of weight -2) the inverse of $\tilde{\tilde{h}}^{ab}$, we obtain the important relation
			\begin{equation}
				q^{I}{}_{J}:=h_{a b}\tilde{B}^{a I}\tilde{B}^{b}{}_{J}=\delta^{I}_{J} - \tau m^{I}m_{J},
			\end{equation}
		which is the projector onto the orthogonal plane to $m^I$ in the internal Minkowski space.
			
		Before addressing the remaining second-class constraints, let us introduce the covariant derivate compatible with $\tilde{B}^{a I}$ that satisfies
			\begin{equation}
				\label{covariantderivate}
				\nabla_{a}\tilde{B}^{b I}:=\partial_{a}\tilde{B}^{b I} + \Gamma_{a}{}^{I}{}_{J}\tilde{B}^{b J} + \Gamma^{b}{}_{a c}\tilde{B}^{c I} - \Gamma^{c}{}_{a c}\tilde{B}^{b I}=0,
			\end{equation}
		where $\Gamma_{a I J}=-\Gamma_{a J I}$ and $\Gamma^{a}{}_{b c}=\Gamma^{a}{}_{c b}$. The 36 equations in Eq.~\eqref{covariantderivate} allow us to completely determinate $\Gamma_{aIJ}$ and $\Gamma^{a}{}_{bc}$. In fact, we find that $\Gamma^{a}{}_{b c}$ are the Christoffel symbols for the three-dimensional metric $q_{ab}$, whereas
			\begin{eqnarray}
				\Gamma_{aIJ} & = &h_{ab}\tilde{B}^{c}{}_{[I|}\partial_{c} \tilde{B}^{b}{}_{|J]} + h_{ab}h_{cd}\tilde{B}^{c}{}_K \tilde{B}^{b}{}_{[I}\tilde{B}^{f}{}_{J]}\partial_{f} \tilde{B}^{dK}  \notag \\ 
					& &+  h_{bc} \tilde{B}^{b}{}_{[I|}\partial_{a} \tilde{B}^c{}_{|J]} - h_{ab}h_{cd} \tilde{B}^{b}{}_K \tilde{B}^{c}{}_{[I}\tilde{B}^{f}{}_{J]} \partial_{f} \tilde{B}^{dK}  \notag \\
					& & - \tau h_{ab} \tilde{B}^{c}{}_{[I}m_{J]} m_{K} \partial_{c} \tilde{B}^{bK}\notag \\
					& &  + \tau h_{bc}\tilde{B}^{b}{}_{[I}m_{J]}m_{K} \partial_{a} \tilde{B}^{cK}.
				\label{GaIJ}	
			\end{eqnarray}	
					
			To guarantee the solution of Eq.~\eqref{Psi} in terms of canonical variables, we pay attention to the kinetic term of Eq. \eqref{Holst} and note that Eq. \eqref{PiSol} implies the following reduction of the symplectic potential:
			\begin{equation}
				\label{sym.estr}
				\ovg{\tilde{\Pi}}{^{a I J}}\partial_{3}{\omega_{a I J}}=\tilde{B}^{a I}\partial_{3}{C_{a I}},
			\end{equation}
			where we have defined
			\begin{equation}
				\label{CaI}
				C_{a I}:=\epsilon\left( \ovg{\omega}_{a I J}m^{J} + m_{I}\ovg{\omega}_{b J K}h_{a c}\tilde{B}^{c J}\tilde{B}^{b K}\right).
			\end{equation}
			The 12 variables $C_{a I}$ can be regarded as the independent components of the connection $\omega_{a I J}$ that are left after the implementation of the symplectic reduction entailed by Eq.~\eqref{PiSol}. The components not showing up in the symplectic structure thus correspond to the ones fixed through the solution of Eq.~\eqref{Psi}.
			
			To solve the constraint \eqref{Psi}, we first observe that it defines a system of six inhomogeneous linear equations for the 18 unknowns of $\omega_{a I J}$. We then parametrize its solution in terms of the $C_{a I}$ introduced above as
			\begin{equation}
			\label{omega}
			\omega_{a I J}= M_{a}{}^{b}{}_{I J K}C_{b}{}^{K} + N_{a I J}.
			\end{equation}
			The first term on the right corresponds to the homogeneous solution of Eq. \eqref{Psi}, whereas the second term designates the particular solution (which is related to the components of the connection that get fixed by the solution of the second-class constraints). We find that $M_{a}{}^{b}{}_{I J K}$ and $N_{a I J}$ are given explicitly as follows:
			\begin{eqnarray}
				\label{M}
				M_{a}{}^{b}{}_{I J K}&=&\epsilon\tau \mathlarger{\mathlarger{\mathlarger{\mathlarger{[}}}}-\delta^{b}_{a}m_{[I}\eta_{J] K} + \delta^{b}_{a}\left(P^{-1}\right)_{I J K L}m^{L} \notag \\
					&&-\left(P^{-1}\right)_{I J L M}h_{a c}\tilde{B}^{c L}\tilde{B}^{b M}m_{K} \notag \\
					&&\left. +\frac{1}{\gamma}\star \left(P^{-1}\right)_{I J L M}h_{a c}\tilde{B}^{b L}m^{M}\tilde{B}^{c}{}_{K}\right], \\
				\label{N}
				N_{a I J}&=& -\tau\underaccent{\tilde}{\lambda}_{a b}\left( \epsilon_{I J K L}\tilde{B}^{b K}m^{L} + \frac{2}{\gamma}\tilde{B}^{b}{}_{[I}m_{J]}
				\right),\quad
			\end{eqnarray}
			where $(P^{-1})^{IJ}{}_{KL}$ is the inverse of $P^{IJ}{}_{KL}$, namely $P^{IJ}{}_{MN} (P^{-1})^{MN}{}_{KL} = \delta^{I}_{[K}\delta^{J}_{L]}$, and
			\begin{eqnarray}
				\label{lambda}
				\underaccent{\tilde}{\lambda}_{a b}:=&&\frac{1}{2}\epsilon_{I J K L}\left(h_{a b}h_{c d}- 2h_{c(a}h_{b)d}\right)\notag\\
				&&\times\tilde{B}^{c I}\tilde{B}^{d M}\tilde{B}^{f J} m^{L}\Gamma_{f}{}^{K}{}_{M}.
			\end{eqnarray}
			We point out that Eq.~\eqref{omega} can be alternatively expressed as $\omega_{a I J}=\Gamma_{aIJ}+J_{a I J}$, where $J_{a I J}$ contains the whole dependence on $C_{aI}$; in this splitting $\Gamma_{aIJ}$ corresponds to a the particular solution of Eq.~\eqref{Psi}, whereas $J_{a I J}$ encodes the homogeneous solution.
			
			Now that we are left with first-class constraints only, the action \eqref{Holst} acquires the form
			\begin{equation}
				\label{Holst5}
				S = \int_{\mathbb{R}} dx^3\int_{\Sigma}dV \Bigl( \tilde{B}^{aI} \partial_{3}{C_{aI}} - \tilde{H}' \Bigr),
			\end{equation}
		    with $\tilde{H}'$ being the Hamiltonian density made up of the first-class constraints,
			\begin{equation}
			\tilde{H}' :=  \lambda_{IJ} \tilde{\mathcal{G}}^{IJ} + N^{a}\tilde{\mathcal{V}}_{a} + \underaccent{\tilde}{N} \tilde{\tilde{\mathcal{H}}},
			\end{equation}
			which in terms of the new canonical coordinates $(C_{a I}, \tilde{B}^{a I})$ are given by
			\begin{subequations} 
				\label{constraintsC}
				\begin{eqnarray}
					&&\tilde{\mathcal{G}}^{I J} = \tilde{B}^{a[I}C_{a}{}^{J]} + 2\epsilon P^{IJ}{}_{KL}\tilde{B}^{a[M}m^{K]}\Gamma_{a}{}^{L}{}_{M} \approx 0, \label{gaussCB} \\
					&&\tilde{\mathcal{V}}_{a} = \nabla_{[b}(\tilde{B}^{bI}C_{a]I}) + \epsilon\tilde{B}^{b[I}m^{K]}\ovg{\Gamma}_{a I J}\Gamma_{b}{}^{J}{}_{K} \nonumber\\
						&&\hspace{8mm}-\tau\tilde{\mathcal{G}}^{I J}(\epsilon C_{aI} - \ovg{\Gamma}_{a I K}m^{K})m_{J}\approx 0,\label{vecCB}\\
					&&\tilde{\tilde{\mathcal{H}}} = -\frac{\tau}{8}\tilde{B}^{a I}\tilde{B}^{b J}R_{a b I J} \nonumber\\
						&&\hspace{1mm}+\frac{1}{4}\tilde{B}^{a [I|}\tilde{B}^{b|J]}\left[C_{a I}C_{bJ} - 2\epsilon C_{a I}\ovg{\Gamma}_{b J K}m^{K}\right. \nonumber\\
						&&\hspace{1mm}+ \left. \left( \Gamma_{a I L} + \frac{2}{\gamma} \star \Gamma_{a I L} \right)\Gamma_{b J K}m^{K}m^{L} - \frac{\tau}{\gamma^{2}}q^{K L}\Gamma_{a I K}\Gamma_{b J L}\right] \nonumber\\
						&&\hspace{1mm} -\frac{\epsilon}{2}\tilde{B}^{aI}m^{J}\nabla_{a}\tilde{\mathcal{G}}_{I J} + \frac{\tau\Lambda}{8}\sqrt{|h|}\approx 0\label{scalCB},
				\end{eqnarray}
			\end{subequations}
		
			\noindent
			where the terms proportional to $\tilde{\mathcal{G}}^{I J}$ squared have been neglected and  $R_{a b I J}:=2(\partial_{[a}{\Gamma_{b]IJ}} + \Gamma_{[a|I}{}^{K}\Gamma_{|b]KJ})$ denotes the curvature of $\Gamma_{a I J}$. Note that $\tau$ shows up in Eqs. \eqref{vecCB} and \eqref{scalCB}, although the term where it appears in the former is proportional to the Gauss constraint. On the other hand, in the latter the value of $\tau$ changes the sign of the term proportional to $R_{a b I J}$ and the sign of the cosmological term (the last term inside the square brackets also depends on $\tau$, but this term can be absorbed through a canonical transformation; see below), indicating that spacelike and timelike foliations have associated different scalar constraints.

		\subsection{Other manifestly Lorentz-covariant phase-space variables}

			As noted in Ref.~\cite{Montesinos1801}, there exist different manifestly Lorentz-covariant parametrizations of the phase space of general relativity related to one another by canonical transformations. These canonical transformations can also be implemented in the present setting. To that end, we consider the phase-space maps $(C_{a I}, \tilde{B}^{a I})\mapsto(K_{a I}, \tilde{B}^{a I})$ and $(C_{a I}, \tilde{B}^{a I})\mapsto(Q_{a I}, \tilde{B}^{a I})$, where $C_{aI}$, $K_{aI}$ and $Q_{aI}$ are related among them by 
			\begin{subequations}
				\label{canonical transformations}
				\begin{eqnarray}
					C_{a I}&=&K_{a I} + \epsilon \left(\Gamma_{aIJ}m^{J} + h_{a b}\tilde{B}^{b J}\tilde{B}^{c K}\Gamma_{c J K}m_{I}\right),\label{transfCK}\\
					\label{cantransformationQ}C_{a I}&=&Q_{a I} + \epsilon \left(\ovg{\Gamma}_{a I J}m^{J} + h_{a b}\tilde{B}^{b J}\tilde{B}^{c K}\ovg{\Gamma}_{c J K}m_{I}\right). \nonumber\\ \label{transfCQ}
				\end{eqnarray}
			\end{subequations}
			These maps are truly canonical transformations, since the symplectic term of the action \eqref{Holst5} changes in each case by a boundary term,
			 \begin{subequations}\label{canvar}
			 	\begin{eqnarray}
			 	\int_{\Sigma}dV\tilde{B}^{aI}\partial_3 C_{aI}\label{canvar1}	&&=\int_{\Sigma}dV\left[ \tilde{B}^{aI}\partial_3 K_{aI} +\partial_a\left(\epsilon m_I\partial_3 \tilde{B}{}^{aI}\right)\right],\nonumber\\
			 	\label{canvar2}\\
			 	\int_{\Sigma}dV\tilde{B}^{aI}\partial_3 C_{aI}&&=\int_{\Sigma}dV\biggl[\tilde{B}^{aI}\partial_3Q_{aI}+\partial_a\biggl(\epsilon m_I\partial_3\tilde{B}{}^{aI}\nonumber\\
			 	&&\hspace{2mm}-\frac{\epsilon\tau}{2\gamma}\sqrt{|h|}\tilde{\eta}^{abc}h_{bd}h_{ce}\partial_3\tilde{B}^{dI}\tilde{B}^e{}_I\biggr)\biggr].\label{canvar3}
			 	\end{eqnarray}
			 \end{subequations}
            In terms of the canonical variables $(K_{a I}, \tilde{B}^{a I})$ the constraints \eqref{gaussCB}--\eqref{scalCB} read
			\begin{subequations}
				\label{constraintsK}
				\begin{eqnarray}
					&&\tilde{\mathcal{G}}^{I J}=\tilde{B}^{a[I}K_{a}{}^{J]} + \frac{\epsilon}{\gamma}\epsilon^{I J}{}_{KL}\tilde{B}^{a[M}m^{K]}\Gamma_{a}{}^{L}{}_{M}\approx 0,\quad  \\
					&&\tilde{\mathcal{V}}_{a}=\nabla_{[b}\left( \tilde{B}^{b I}K_{a] I}\right) + \frac{\epsilon}{\gamma}\tilde{B}^{b[I}m^{K]}\star\Gamma_{a I J}\Gamma_{b}{}^{J}{}_{K} \nonumber\\
					&&\hspace{9mm}-\tau\tilde{\mathcal{G}}^{I J}\left(\epsilon K_{aI} - \frac{1}{\gamma}\star \Gamma_{a I K}m^{K}\right) m_{J} \approx 0,\\
					&&\tilde{\tilde{\mathcal{H}}}= -\frac{\tau}{8}\tilde{B}^{a I}\tilde{B}^{b J}R_{a b I J} \nonumber \\
					&&\hspace{4mm}+ \frac{1}{4}\tilde{B}^{a[I|}\tilde{B}^{b|J]}\left[K_{aI}K_{bJ} - \frac{2\epsilon}{\gamma}K_{aI}\star \Gamma_{b J K}m^{K}\right. \nonumber\\
					&&\hspace{4mm} \left. -\frac{\tau}{\gamma^{2}}q^{K L}\Gamma_{a I K}\Gamma_{b J L}\right] - \frac{\epsilon}{2}\tilde{B}^{aI}m^{J}\nabla_{a}\tilde{\mathcal{G}}_{I J} \nonumber \\
					&&\hspace{4mm}+\frac{\tau\Lambda}{8}\sqrt{|h|}\approx 0.
				\end{eqnarray}
			\end{subequations}
		Likewise, for the canonical pair $(Q_{a I}, \tilde{B}^{a I})$ we obtain
			\begin{subequations}
				\label{constraintsQ}
				\begin{eqnarray}
					&&\tilde{\mathcal{G}}^{I J} =\tilde{B}^{a [I}Q_{a}{}^{J]} \approx 0, \label{GaussQ}\\
					&&\tilde{\mathcal{V}}_{a} =\nabla_{[b}\left(\tilde{B}^{bI}Q_{a]I}\right) -\epsilon\tau Q_{a I}m_{J}\tilde{\mathcal{G}}^{I J} \approx 0,\label{VectQ}\\
					&&\tilde{\tilde{\mathcal{H}}}=-\frac{\tau}{8}\tilde{B}^{a I}\tilde{B}^{b J}R_{a b I J} + \frac{1}{4}\tilde{B}^{a[I|}\tilde{B}^{b|J]}Q_{a I}Q_{bJ} \nonumber\\
					&&\hspace{8mm}-\frac{\epsilon}{2}\tilde{B}^{a I}m^{J}\nabla_{a}\tilde{\mathcal{G}}_{I J} + \frac{\tau\Lambda}{8}\sqrt{|h|}\approx 0.\label{ScalQ}
				\end{eqnarray}
			\end{subequations}
			In all cases, the diffeomorphism constraint reads
			\begin{equation}\label{diffeomorphism}
				\tilde{\mathcal{D}}_{a}=\tilde{B}^{bI}\partial_{[b}U_{a]I} + \frac{1}{2}U_{aI}\partial_{b}\tilde{B}^{bI},
			\end{equation}
			for $U_{a I}=C_{a I}, K_{aI}$ or $Q_{aI}$. As expected, our results with $\tau=-1$ reproduce those of Ref.~\cite{Montesinos1801} with $\sigma=-1$. It is worth stressing that the constraints \eqref{GaussQ}--\eqref{ScalQ} correspond to those resulting in the canonical analysis of the Palatini action, and thus the canonical transformation involving Eq.~\eqref{transfCQ} links the Hamiltonian formulations of the Palatini and Holst actions.

			
	\section{\label{sec:GF} The Space Gauge}
		
		As mentioned above, the space gauge reduces the internal group $SO(1,3)$ to its subgroup $SU(1,1)$. In the present framework, we can achieve this by taking $\tau=1$ and imposing the gauge condition\footnote{Any of the internal spatial directions can be equivalently used to perform the gauge fixing.} $\tilde{B}^{a3}=0$, which amounts to $m^{i}=0$ provided that $\det(\tilde{B}^{ai}) \neq 0$ (assumed throughout this section). Notice that both conditions imply, from the normalization of $m_{I}$, that $(m^{3})^{2}=1$; this is why we consider $\tau=1$, since $\tau=-1$ would make complex the ensuing formulation with respect to a spacelike direction (the case $\tau=-1$ works for the gauge condition $\tilde{B}^{a0}=0$, as shown in Ref.~\cite{Montesinos1801}). On the other side, this gauge condition Poisson commutes (modulo the gauge condition itself) with all the constraints except for $\tilde{\mathcal{G}}^{i 3}$ (this holds for any set of canonical variables), which results in
		\begin{equation}
			\label{PB1}
			\left\lbrace \tilde{B}^{a3}(x), \tilde{\mathcal{G}}^{i3}(y)\right\rbrace = -\frac{1}{2}\tilde{B}^{ai}\delta^{3}(x,y).
		\end{equation}
		Since we assumed that $\tilde{B}^{ai}$ is nonsingular, the set $(\tilde{B}^{a3},\tilde{\mathcal{G}}^{i3})$ is second class, and so the constraint $\tilde{\mathcal{G}}^{i3}$ (one boost and two rotations) must be simultaneously solved. Its solution reads $C_{a3} = \epsilon m_{3}\tilde{B}^{bi}\partial_{b}\underaccent{\tilde}{B}_{ai}$ (or $K_{a3}=Q_{a3}=0$ for the others parametrizations), where $\underaccent{\tilde}{B}_{ai}$ denotes the inverse of $\tilde{B}^{ai}$. Meanwhile, by defining $\epsilon_{i j k}:=\epsilon_{i j k 3}$ and $\tilde{\mathcal{G}}_{i} := -\frac{1}{2} \epsilon_{ijk} \tilde{\mathcal{G}}^{j k}$, and noticing that the internal metric is now $\big(\eta_{ij}\big) = \mbox{diag}(-1,1,1)$ (the induced metric on the hypersurface $\Sigma$ is $q^{ab}=2|\det(\tilde{B}^{ck})|^{-1}\eta_{ij}\tilde{B}^{ai}\tilde{B}^{bj}$), the remaining Gauss constraints satisfy the algebra
		\begin{equation}\label{algebraGauss}
			\left\lbrace \tilde{\mathcal{G}}_{i}(x), \tilde{\mathcal{G}}_{j}(y) \right\rbrace = \frac{1}{2}\epsilon_{ij}{}^{k}\tilde{\mathcal{G}}_{k}
			\delta^{3}(x,y),
		\end{equation}
		which corresponds to the Lie algebra of the group $SU(1,1)$, whose members are one rotation ($\tilde{\mathcal{G}}_{0}$) and two boost generators ($\tilde{\mathcal{G}}_{1}$ and $\tilde{\mathcal{G}}_{2}$). Therefore, as promised, the above gauge breaks the four-dimensional Lorentz group $SO(1,3)$ down to its three-dimensional counterpart $SU(1,1)$.
		
		Regarding the connection $\Gamma_{aIJ}$, under the gauge fixing we obtain, from Eq.~\eqref{GaIJ}, that $\Gamma_{ai3}=0$, whereas $\Gamma_{a i}:= -(1/2)\epsilon_{i j k}\Gamma_{a}{}^{j k}$ happens to be the spin connection compatible with $\tilde{B}^{ai}$,
		\begin{equation}
			\label{connectionGauge}
			\Gamma_{ai} = -\epsilon_{i j k}\left(\partial_{[b}\underaccent{\tilde}{B}_{a]}{}^{j} + \underaccent{\tilde}{B}_{a}{}^{[l|}\tilde{B}^{c|j]}\partial_{b}
			\underaccent{\tilde}{B}_{cl}\right)\tilde{B}^{bk}.
		\end{equation}
		
		Before moving on to expressing the constraints in the space gauge, let us see how the canonical transformations \eqref{transfCK}--\eqref{transfCQ} are affected by it. Setting $\tilde{B}^{a3}=0$ there, we obtain
		\begin{subequations}
			\begin{eqnarray}
				\label{Ca3} C_{a3}&=& \epsilon m_{3}\tilde{B}^{bi}\partial_{b}\underaccent{\tilde}{B}_{ai},\\
				\label{Cai} 
				C_{ai}&=&K_{ai}=Q_{ai} - \frac{\epsilon}{\gamma}m^{3}\Gamma_{ai}.
			\end{eqnarray}
		\end{subequations}
		The first line gives no new information, but after defining $A_{ai}:=-\gamma \epsilon m^{3}C_{ai}$, the second line becomes the analog of Barbero's canonical transformation
		\begin{equation}
			\label{Aconnection}
			A_{ai}= -\gamma \epsilon m^{3} Q_{ai} + \Gamma_{ai}.
		\end{equation} 
		Since $\Gamma_{a i}$ is an $SU(1,1)$ connection and $Q_{a i}$ is an internal vector, $A_{a i}$ is  an $SU(1,1)$ connection as well.
		
		Let us consider the formulation described by the canonical variables $(C_{ai},\tilde{B}^{ai})$ or equivalently $(K_{ai},\tilde{B}^{ai})$. Observe that the kinetic term in Eq.~\eqref{Holst5} can be expressed as
		\begin{equation}
			\tilde{B}^{aI}\partial_{3}{C_{aI}}=\tilde{B}^{ai}\partial_{3}{C_{ai}}=\frac{2}{\gamma}\tilde{E}^{ai} \partial_{3}{A_{ai}},
		\end{equation}
		where the first equality is a consequence of the gauge fixing, and for the second equality to hold we have defined
		\begin{equation}
			\label{triad}
			\tilde{E}^{a i}:=-\frac{1}{2}\epsilon m^{3}\tilde{B}^{ai}.
		\end{equation}
		Thus,  $A_{a i}$ and $\tilde{E}^{a i}$ are canonically conjugate to each other and satisfy the fundamental Poisson bracket $\left\lbrace A_{ai}(x), \tilde{E}^{bj}(y) \right\rbrace = (\gamma /2)\delta_{a}^{b}\delta_{i}^{j} \delta^{3}(x,y)$. Using these variables, the first-class constraints \eqref{gaussCB}--\eqref{scalCB} after the gauge fixing read
		\begin{subequations} \label{Barbero}
			\begin{eqnarray}
				\tilde{\mathcal{G}}_{i}&=&-\frac{1}{\gamma}\left( \partial_{a}\tilde{E}^{a}{}_{i} - \epsilon_{i j k}A_{a}{}^{j}\tilde{E}^{a k}\right) \approx 0, \label{BarbGauss}\\
				\tilde{\mathcal{V}}_{a}&=&\frac{1}{\gamma}\tilde{E}^{b i}F_{ba i} - (A_{a i}- \Gamma_{a i})\tilde{\mathcal{G}}^{i}\approx 0,\label{BarbVect}\\
				\tilde{\tilde{\mathcal{H}}}&=& \frac{1}{2 \gamma ^{2}} \epsilon_{i j k} \tilde{E}^{a i} \tilde{E}^{b j} \left[ F_{ab}{}^{k} - (\gamma^{2} + 1)R_{ab}{}^{k}\right] \nonumber \\
				&& + \frac{1}{\gamma} \tilde{E}^{a i} \nabla_{a}\tilde{\mathcal{G}}_{i} + \Lambda |\tilde{\tilde{E}}| \approx 0,\label{BarbScal}
			\end{eqnarray}
		\end{subequations}
		where $\tilde{\tilde{E}}=\det(\tilde{E}^{a i})$, $\Gamma_{ai}$ is given by the same expression \eqref{connectionGauge} with the change $B\rightarrow E$ (of course, $\underaccent{\tilde}{E}_{ai}$ is the inverse of $\tilde{E}^{ai}$), and $F_{ab i}$ and $R_{abi}$ are the respective curvatures of $A_{a i}$ and $\Gamma_{ai}$,
		\begin{subequations}
		\begin{eqnarray}
			F_{abi} & := & 2\partial_{[a}A_{b]i} - \epsilon_{i j k}A_{a}{}^{j}A_{b}{}^{k},\label{Fabi}\\
			R_{abi} & := & -\frac{1}{2}\epsilon_{ijk}R_{ab}{}^{jk}=2\partial_{[a}\Gamma_{b]i} - \epsilon_{i j k}\Gamma_{a}{}^{j}\Gamma_{b}{}^{k}.\quad\label{Rabi}
		\end{eqnarray}
		\end{subequations}
		The formulation embodied in the set of constraints \eqref{BarbGauss}--\eqref{BarbScal} has exactly the same form as Barbero's one, but they differ in the internal gauge group; while the latter makes use of the time gauge to reduce the Lorentz group to $SO(3)$, the former employs the space gauge in order to single out $SU(1,1)$ as the residual internal symmetry. An alternative way of obtaining Eqs.~\eqref{BarbGauss}-\eqref{BarbScal} (although less directly) is analogous to the path originally followed by Barbero to obtain his variables, which consists in using the canonical transformation \eqref{Aconnection} in the Hamiltonian formulation arising from the Palatini action subject to the same gauge fixing. As mentioned at the end of Sec.~\ref{sec:FCH}, that Hamiltonian formulation is equivalent to the one that employs the canonical pair $(Q_{aI},\tilde{B}^{aI})$. Hence, for the sake of completeness, we display here the constraints \eqref{GaussQ}--\eqref{ScalQ} after imposing the space gauge,
		\begin{subequations}\label{GaugeQ}
			\begin{eqnarray}
			\tilde{\mathcal{G}}_{i}&=&-\frac{1}{2}\epsilon_{i j k} \tilde{B}^{aj}Q_{a}{}^{k}\approx 0, \\
			\tilde{\mathcal{V}}_{a}&=&\nabla_{[b}\left( \tilde{B}^{bi}Q_{a]i}\right) \approx 0, \\
			\tilde{\tilde{\mathcal{H}}}&=&-\frac{1}{8}\epsilon_{i j k}\tilde{B}^{ai}\tilde{B}^{bj}R_{ab}{}^{k} + \frac{1}{4}\tilde{B}^{a[i|}\tilde{B}^{b|j]}
			Q_{ai}Q_{bj} \nonumber\\
			&&+\frac{\Lambda}{8}\sqrt{|h|} \approx 0,
			\end{eqnarray}
		\end{subequations}
		where the expression for $\Gamma_{ai}$ is understood to be taken from Eq.~\eqref{connectionGauge}. If we wanted to put a name to this formulation, it might be called the $SO(1,2)$ ADM formalism, in analogy with the $SO(3)$ case~\cite{AshtLectures,AshBalJo}.
		
		\section{\label{sec:AA} Alternative derivation of the $SU(1,1)$ Barbero-like variables}
			
			To enrich the content of this work, in this section we show how to obtain the canonical formulation \eqref{BarbGauss}--\eqref{BarbScal} starting from a different parametrization of the solution of the second-class constraints, which can also be solved in a nonexplicitly Lorentz-covariant fashion (see Appendix). The approach of this section is thus closer to that of Ref.~\cite{Barros0100} (see also Ref.~\cite{Romero2016}).
						
			According to the Appendix, instead of parametrizing the solution of the second-class constraints \eqref{Phi} by the 12 variables $\tilde{B}^{aI}$, we can employ the nine plus three variables $(\tilde{E}^{ai},l^i)$. The associated canonical variables are no longer manifestly Lorentz covariant, but they encode, at least at the classical level, exactly the same information. We point out that in order to facilitate the description of a spacetime foliation with respect to a spacelike direction (the ``$x^3$'' direction) within this formalism, it is much better to solve the second-class constraints with respect to the corresponding internal spacelike direction (as compared with the approach of Ref.~\cite{Liu1706}). In these variables, the space gauge corresponds to the gauge condition 
			\begin{equation}
				l^i=0,
			\end{equation}
			which can be readily seen to form a second-class set together with the constraint \eqref{Gauss21}. Indeed, the Poisson bracket between them yields
			\begin{equation}
				\left\{l^i (x),\tilde{\mathcal{G}}^{j3}(y)\right\}\Bigl|_{l^i=0}=-\frac{1}{\gamma}\eta^{ij}\delta^{3}(x,y),
			\end{equation}
			which defines a nonsingular matrix. Thus, as in Sec.~\ref{sec:GF}, the remanent internal symmetry is just $SU(1,1)$.
					
			According to Eq.~\eqref{sympest}, in the space gauge the symplectic term of the action \eqref{Holst} takes the form
			\begin{equation}
			\int_{\mathbb{R}} dx^3\int_{\Sigma}dV \tilde{\Pi}{}^{aIJ} \partial_{3} \ovg{\omega}_{aIJ} = \frac{2}{\gamma}\int_{\mathbb{R}} dx^3\int_{\Sigma}dV \tilde{E}^{ai} \partial_{3} A_{ai},
			\end{equation}
			where we have defined
			\begin{equation}
			\label{Adef}
			A_{ai} := \gamma \ovg{\omega}_{a3i} = \gamma \omega_{a3i} - \frac{1}{2} \epsilon_{ijk} \omega_{a}{}^{jk}.
			\end{equation}
			So, half of the components of the connection $\omega_{aIJ}$ are encoded in $A_{ai}$ and the other half drop out of the symplectic structure. The latter are precisely the ones fixed by the simultaneous solutions of the second-class constraints \eqref{Psi} and $\tilde{\mathcal{G}}^{i3}=0$. Instead of solving Eq.~\eqref{Gauss21} directly for $\tilde{Z}^{i}$ and substituting the resulting solution in the first-class constraints, it turns out to be more enlightening to go a few steps back in the computations and not to use the explicit solution of Eq.~\eqref{Psi}, but rather combine it together with $\tilde{\mathcal{G}}^{i3}$ into a system of nine equations for the components $\omega_{aij}$ (which are half the components of the connection) as follows:
			\begin{eqnarray}
			\label{Gb}
			\tilde{\mathcal{G}}^{i3} & = & - \partial_{a}{\tilde{E}^{ai}} - \frac{1}{\gamma^2} \epsilon^{i}{}_{jk}A_{a}{}^{j} \tilde{E}^{ak} \notag\\
			&&- \left(1+ \frac{1}{\gamma^{2}}\right) \omega_{a}{}^{i}{}_{j} \tilde{E}^{aj}=0, \\
			\label{PsiGF}
			\Psi^{ab} & = & 2 \epsilon_{ijk}\tilde{E}^{(a|i}\tilde{E}^{cj} \left(\partial_{c}{\tilde{E}^{|b)k}} + \omega_{c}{}^{k}{}_{l}\tilde{E}^{|b)l}\right)=0.
			\end{eqnarray}
			The solution for $\omega_{aij}$ then gives
			\begin{equation}\label{wij}
				\omega_{aij}= \epsilon_{ijk}\Gamma_{a}{}^{k} + \dfrac{\gamma}{\gamma^{2}+1} \underaccent{\tilde}{E}_{a[i}\tilde{\mathcal{G}}_{j]},
			\end{equation}
			where
			\begin{equation}
			\Gamma_{ai}:=  -\epsilon_{ijk}\left(\partial_{[b}\underaccent{\tilde}{E}_{a]}{}^{j} + \underaccent{\tilde}{E}_{a}{}^{[l|}\tilde{E}^{c|j]}\partial_{b}
			\underaccent{\tilde}{E}_{cl}\right)\tilde{E}^{bk}
			\end{equation}
			is the spin connection compatible with the densitized triad\footnote{Note that in this case the induced metric on the hypersurface $\Sigma$ takes the form $q^{ab}=|\tilde{\tilde{E}}|^{-1}\eta_{ij}\tilde{E}^{ai}\tilde{E}^{bj}$, which explicitly exhibits the $SU(1,1)$ invariance.} $\tilde{E}^{ai}$, and $\tilde{\mathcal{G}}^{i}:= -(1/2)\epsilon^{i}{}_{jk}\tilde{\mathcal{G}}^{jk}$ are the generators of the internal $SU(1,1)$ symmetry given by the same expression \eqref{BarbGauss}.
			
			The expression for the remaining components of the connection can be obtained from Eqs. \eqref{Adef} and \eqref{wij} as
			\begin{equation}
				\omega_{a3i} = \frac{1}{\gamma} A_{ai} - \frac{1}{\gamma} \Gamma_{ai} +  \dfrac{1}{2(\gamma^{2}+1)}\epsilon_{ijk} \underaccent{\tilde}{E}_{a}{}^{j}\tilde{\mathcal{G}}^{k}.
			\end{equation}
			
			All that is left is to write the vector and scalar constraints in terms of the canonical variables $(A_{ai},\tilde{E}^{ai})$. It is not difficult to show that they take exactly the same form as in Eqs.~\eqref{BarbVect} and \eqref{BarbScal}, respectively, with the same definitions \eqref{Fabi} and \eqref{Rabi} for the curvatures, but they all expressed in terms of the variables of this section (which actually coincide with those introduced in the previous section). In conclusion, we have obtained the same Barbero-like formulation with internal group $SU(1,1)$ of Sec.~\ref{sec:GF}.
		
	\section{Conclusions}
		
		In this paper, we have exploited the advantages of a manifestly Lorentz-covariant formulation to obtain in a straightforward manner gravity as an $SU(1,1)$ gauge theory. We departed from the Holst action with a cosmological constant on a $M=\Sigma\times\mathbb{R}$ spacetime manifold with full local Lorentz invariance, and performed its $3+1$ decomposition with respect to one of the ``spatial'' directions.~Although it is not the usual foliation, we found that the structure of the resulting canonical theory is exactly the same as the one of the mainstream approach, including the presence of second-class constraints. This is a manifestation of the diffeomorphism invariance of the original theory, wherein the distinction between space and time is not relevant.~As in Ref.~\cite{Montesinos1801}, in Sec.~\ref{sec:FCH} we solved the second-class constraints in a manifestly Lorentz-covariant fashion while keeping in mind the timelike nature of $\Sigma$, which is achieved by introducing an internal spacelike direction with respect to which this solution is expressed and the right character of the metric induced on $\Sigma$ is specified. After that, we displayed several real parametrizations---$(C_{a I}, \tilde{B}^{a I})$, $(K_{a I}, \tilde{B}^{a I})$ and $(Q_{a I}, \tilde{B}^{a I})$---of the phase space that are related among them by canonical transformations such that the momentum variables $\tilde{B}^{aI}$ are held fixed whereas the configuration variables are transformed according to Eqs.~\eqref{transfCK} and \eqref{transfCQ}. Notably, the canonical formulation expressed in terms of the variables $(Q_{a I}, \tilde{B}^{a I})$ lacks the presence of the Immirzi parameter (modulo the Gauss constraint) and actually takes the same form as the one arising from the Palatini action, which makes Eq. \eqref{transfCQ} a canonical transformation between it and the Holst action. It should be mentioned that the issue of the boundary conditions for the above kind of foliation was not addressed in this work, but we hope to carefully treat it elsewhere.
		
		The description of gravity as an $SU(1,1)$ gauge theory immediately emerges after imposing the space gauge on the above formulation, where, in the internal Minkowski space, boosts along one determinate spatial direction and rotations around the two perpendicular spatial directions to it are frozen. As a result, the remaining members of the Lorentz group, namely rotations around the same fixed spatial direction and boosts along its two perpendicular directions, comprise the remanent gauge group $SU(1,1)$. Operationally, the space gauge is imposed through the condition $\tilde{B}^{a3}=0$, and, as shown in Sec.~\ref{sec:GF}, the resulting canonical formulation is parametrized by the components of an internal real connection $A_{ai}$ and their canonically conjugate momenta (the densitized triad $\tilde{E}^{ai}$) both subject to a set of constraints that take the same form as those of the Ashtekar-Barbero formalism, but with internal group $SU(1,1)$ instead of $SU(2)$. For the sake of completeness, we established in Sec.~\ref{sec:AA} that the same canonical formulation also comes out after applying the space gauge to a different parametrization of the solution of the second-class constraints that is related to the formulation without manifest Lorentz covariance contained in the Appendix.
		
		The benefit of a manifestly Lorentz-covariant description is seen in the simplicity that the formulation \eqref{BarbGauss}--\eqref{BarbScal} emerges.~This contrasts with the approach of Ref.~\cite{Liu1706}, where the authors had to reconstruct the form of the constraints in accordance with the gauge condition they found, all this in order to exhibit at the end the explicit covariance  under $SU(1,1)$ of the resulting theory. In fact, they did not even displayed the form of the scalar constraint in terms of their $SU(1,1)$ variables, whereas in our case we neatly arrive at Eq.~\eqref{BarbScal}.
		
		On the other hand, recall that the Holst action can be recast as a constrained $BF$ theory, whose Lorentz-covariant canonical analysis~\cite{CelMont} actually leads to a set of constraints that agrees with Eqs.~\eqref{Gauss}--\eqref{Psi} up to terms proportional to the second-class constraint \eqref{Phi}. Therefore, by explicitly solving the second-class constraints as was done in Sec.~\ref{sec:FCH}, the canonical analyses of the $BF$-type and Holst actions completely agree, meaning that the results contained in this paper as well as those of Ref.~\cite{Montesinos1801} also hold for the canonical analysis of $BF$ gravity with the Immirzi parameter, that is, the canonical formulation \eqref{BarbGauss}--\eqref{BarbScal} can also be derived, in the space gauge, from a $BF$-type action for general relativity.
		
		The similarities between our description and Barbero's one make us believe that our approach could be extended to the quantum world using the same strategy of the loop approach (even if the gauge group is not compact; see Ref.~\cite{Freidel}). For instance, it would be particularly interesting to see how the results for the eigenvalues of the area operator (defined on both timelike and spacelike surfaces) compare to those of Ref.~\cite{Liu1706} as well as those obtained in the context of twisted geometries \cite{Rennert2017}, when using our $SU(1,1)$ Barbero-like variables to construct the holonomies.

	\section*{Acknowledgments}
	
		This work was partially supported by Consejo Nacional de Ciencia y Tecnolog\'{i}a (CONACyT), M\'{e}xico, Grants No. 237004-F and No. 237351. M. C. would like to acknowledge the financial support of Programa para el Desarrollo Profesional Docente, para el Tipo Superior (PRODEP) Grant No. 12313153 (through UAM-I).

	\appendix*
	
	\section{Nonmanifestly Lorentz-covariant parametrization}\label{Appendix}	
	
	Although, as shown in Sec.~\ref{sec:FCH}, the second-class constraints \eqref{Psi}--\eqref{Phi} can be solved in a manifestly Lorentz-covariant fashion, they can also be solved in such a way that the Lorentz symmetry is preserved in a nonexplicitly covariant way~\cite{Barros0100} (just as Maxwell electrodynamics can be described in terms of electric and magnetic fields instead of the 4-potential without spoiling Lorentz invariance). Inspired by the work of Ref.~\cite{Barros0100}, we devote this Appendix to do the latter but this time adapting it to a decomposition with respect to spacelike direction.

	We take as our starting point the constraint \eqref{Psi}, whose solution is given by
	\begin{subequations}
	\begin{eqnarray}
		\label{pi1}
		\tilde{\Pi}^{a3i} & = &\tilde{E}^{ai},\label{PhiSol21} \\
		\label{pi2}
		\tilde{\Pi}^{aij} & = & 2\tilde{E}^{a[i}l^{j]},\label{PhiSol22}
	\end{eqnarray}
	\end{subequations}
	 with $l^{i}$ being three arbitrary functions and $\tilde{E}^{ai}$ assumed to have an inverse that we denote by $\underaccent{\tilde}{E}_{ai}$. This solution can be used to rewrite Eq.~\eqref{qPi} as
	\begin{equation}
	\tilde{\tilde{q}}q^{ab} = \Theta_{ij} \tilde{E}^{ai} \tilde{E}^{bj},
	\end{equation}
	for
	\begin{equation}
	\Theta_{ij} := - \left(1+l_{k}l^{k}\right)\eta_{ij} + l_{i}l_{j},
	\end{equation}
	with $(\eta_{ij}) = \mbox{diag}(-1,1,1)$. Moreover, the solution \eqref{PhiSol21}--\eqref{PhiSol22} induces the following symplectic reduction in the action~\eqref{Holst}:	
	\begin{eqnarray}
	\ovg{\tilde{\Pi}}{}^{aIJ} \partial_{3}{\omega_{aIJ}}&=&\tilde{\Pi}{}^{aIJ} \partial_{3} \ovg{\omega}_{aIJ}\notag\\
	& = &\frac{2}{\gamma}  \left(  \tilde{E}^{ai} \partial_{3} A_{ai}  + \tilde{Z}^{i}\partial_{3}l_{i}\right),\label{sympest}
	\end{eqnarray}
	where the definitions
	\begin{eqnarray}
	\label{Adef2}
	A_{ai} & := & \gamma \ovg{\omega}_{a3i} + \gamma \ovg{\omega}_{aij}l^{j}, \\
	\label{Zdef}
	\tilde{Z}_{i} & := & \gamma \ovg{\omega}_{aij}\tilde{E}^{aj},
	\end{eqnarray}
	provide a set of 12 relations that, together with the six remaining second-class constraints \eqref{Psi}, allow us to completely express the 18 components of $\ovg{\omega}_{aIJ}$ (or equivalently of $\omega_{aIJ}$) in terms of the new canonical variables $\left(A_{ai},\tilde{E}^{ai}\right)$ and $\left(l_{i},\tilde{Z}^{i}\right)$. Equations \eqref{Adef2} and \eqref{Zdef} imply that we can parametrize the components of the connection as
	\begin{subequations}
		\label{og}
		\begin{eqnarray}
		\label{og1}
		\ovg{\omega}_{a3i}  &=&  \frac{1}{\gamma} A_{ai} - \frac{1}{2}\epsilon_{ijk}\underaccent{\tilde}{E}_{al} l^{j}\tilde{M}^{kl} + \frac{1}{\gamma} \underaccent{\tilde}{E}_{a[i}\tilde{Z}_{j]}l^{j},\quad \\
		\ovg{\omega}_{aij}  &=& \frac{1}{2}\epsilon_{ijk}\underaccent{\tilde}{E}_{al} \tilde{M}^{kl} - \dfrac{1}{\gamma}\underaccent{\tilde}{E}_{a[i}\tilde{Z}_{j]}, \label{og2}
		\end{eqnarray}
	\end{subequations}
	where $\tilde{M}^{ij} = \tilde{M}^{ji}$ are six independent parameters that will be fixed by the constraints \eqref{Psi}. Before going further, note that the Gauss constraint in terms of the new canonical variables takes the form
	\begin{subequations}
		\label{Gauss2}
		\begin{eqnarray}
		\tilde{\mathcal{G}}^{i3} =&&  - \partial_{a} \left(\tilde{E}^{ai} - \frac{1}{\gamma}\epsilon^{i}{}_{jk}\tilde{E}^{aj}l^{k} \right) + \frac{2}{\gamma} \tilde{E}^{a[i}l^{j]}A_{aj}\notag \\
		&& - \frac{1}{\gamma} \left(\tilde{Z}^{i} + l^{i}l^{j}\tilde{Z}_{j} \right),\label{Gauss21}\\
		\tilde{\mathcal{G}}^{i} :=&& -\frac{1}{2} \epsilon^{i}{}_{jk} \tilde{\mathcal{G}}^{jk} = - \partial_{a}\left( \frac{1}{\gamma} \tilde{E}^{ai} + \epsilon^{i}{}_{jk}\tilde{E}^{aj}l^{k} \right)  \notag \\
		& & - \frac{1}{\gamma} \epsilon^{i}{}_{jk} \left( \tilde{E}^{aj}A_{a}{}^{k} + \tilde{Z}^{j}l^{k}\right),\label{Gauss22}
		\end{eqnarray}
	\end{subequations}
	which demonstrates how the nature of the solution \eqref{PhiSol21}--\eqref{PhiSol22} compels us to split the manifestly covariant form of the Gauss constraint into $\tilde{\mathcal{G}}^{i3}$ and $\tilde{\mathcal{G}}^{jk}$. The splitting does not break Lorentz covariance though, since $\tilde{\mathcal{G}}^{i3}$ and $\tilde{\mathcal{G}}^{i}$ still generate the whole local Lorentz symmetry. Moving on to the solution of Eq.~\eqref{Psi}, we substitute Eqs. \eqref{PhiSol21}--\eqref{PhiSol22} and \eqref{og1}--\eqref{og2} in it, and multiplying that result by $\underaccent{\tilde}{E}_{a i} \underaccent{\tilde}{E}_{b j}$, we obtain the equation
	\begin{equation}
	\label{PsiReduce}
	\tilde{F}_{(ij)} - \left(1+l_{p} l^{p} \right)^{2} \epsilon_{ikm}\epsilon_{jln} \left( \Theta^{-1} \right)^{m n} \tilde{M}^{k l}=0,
	\end{equation}
	where $\left(\Theta^{-1}\right)_{ij}$, the inverse of $\Theta_{ij}$, and $\tilde{F}_{ij}$ are given respectively by
	\begin{eqnarray}
	&&\left(\Theta^{-1}\right)_{ij} =   - \dfrac{1}{1+l_{k} l^{k}} \left(\eta_{ij} + l_{i}l_{j}\right),\\
	&&\tilde{F}_{ij}  =   \left( 1+l_{k}l^{k} \right) \tilde{f}_{ij} + 
	\left(\eta_{ik} -\frac{1}{\gamma}\epsilon_{ikl}l^{l}\right)l_{j}\tilde{\mathcal{G}}^{k}  \notag\\
	\label{F}
	&&\hspace{9mm}- \left(\frac{1}{\gamma} \eta_{ik} + \epsilon_{ikl}l^{l}\right) l_{j}\tilde{\mathcal{G}}^{k3}, 
	\end{eqnarray}
	for
	\begin{eqnarray}
	\label{f}
	\tilde{f}_{ij} & = & \epsilon_{ikl}\tilde{E}^{ak}\left[\left(1+\frac{1}{\gamma^{2}}\right) \underaccent{\tilde}{E}_{bj} \partial_{a}\tilde{E}^{bl} + \frac{1}{\gamma} A_{aj} l^{l} \right] \notag \\ 
	& & -\frac{1}{\gamma^2}\left(\tilde{E}^{ak}A_{ak}\eta_{ij} - \tilde{E}^{a}{}_{i}A_{aj}+l_{i}\tilde{Z}_{j} \right). 
	\end{eqnarray}
	Solving  Eq.~\eqref{PsiReduce} for $M_{ij}$ we obtain
	\begin{eqnarray}
	\tilde{M}_{i j} =& & \frac{2}{(1+l_{r}l^{r})^{2}} \left[ \delta^k_i \delta^l_j - \frac{1}{4} \left( \Theta^{-1} \right)_{ij} \Theta^{k l}\right]\notag\\
	&&\times \epsilon_{k m p} \epsilon_{l n q}\Theta^{m n}\tilde{F}^{(p q)},\\
	= &&  \frac{1}{(1+ l_{m}l^{m})^{2}}\Big[ - 2 \tilde{F}_{(ij)} + \left(\tilde{F}^{k}{}_{k}+ \tilde{F}_{kl}l^{k}l^{l} \right)\eta_{ij}   \notag \\
	\label{Mij}
	& &+ \left(\tilde{F}^{k}{}_{k} - \tilde{F}_{kl} l^{k}l^{l} \right)l_{i}l_{j} -2 \left(l_{(i}\tilde{F}_{j)k} + \tilde{F}_{k(i}l_{j)}\right)l^{k}  \Big].\notag\\
	\end{eqnarray}
	
	Hence $\ovg{\omega}_{aIJ}$ is completely determined by Eqs. \eqref{og1}--\eqref{og2}, \eqref{F}, \eqref{Mij} and \eqref{f}; using the resulting expressions together with the solution \eqref{PhiSol21}--\eqref{PhiSol22}, the vector and scalar  constraints read
	\begin{eqnarray}
	\tilde{\mathcal{V}}_{a} & = & \dfrac{2}{\gamma} \tilde{E}^{bi} \partial_{[b}{A_{a]i}} - \frac{1}{\gamma} \tilde{Z}_{i}\partial_{a}{l^{i}} + \dfrac{1}{\gamma^{2}+1}\Bigg[A_{ai}\Big( 2\tilde{E}^{b[i}l^{j]}A_{bj} \notag \\
	& &\hspace{-5mm} - \tilde{Z}^{i} - \tilde{Z}_{j}l^{j}l^{i} \Big) +\dfrac{1}{\gamma} \epsilon_{ijk} A_{a}{}^{i}\left(A_{b}{}^{j}\tilde{E}^{bk}+ l^{j}\tilde{Z}^{k}\right)\Bigg],\\
	\tilde{\tilde{\mathcal{H}}} & = & \tilde{E}^{ai}l_{i}\tilde{\mathcal{V}}_{a} \notag\\
	&& \hspace{-5mm}- \frac{1}{\gamma} \left(1+ l_{k}l^{k}\right)\left(  \tilde{E}^{ai}\partial_{a}{\tilde{Z}_{i}} + \frac{1}{2} \tilde{E}^{ai}\tilde{E}^{bj}\tilde{Z}_{i}\partial_{a}{\underaccent{\tilde}{E}_{bj}}\right) \notag \\
	& &\hspace{-5mm} - \dfrac{\gamma^2}{\gamma^{2}+1} \Bigg\lbrace \frac{1}{\gamma^2}\left(1+l_{l}l^{l}\right)\Bigg[ - \frac{3}{4} \tilde{Z}_{i}\tilde{Z}^{i} - \frac{3}{4} \left(\tilde{Z}_{i}l^{i} \right)^{2} \notag \\
	& &\hspace{-5mm} -\tilde{E}^{ai} A_{ai}\tilde{Z}^{j}l_{j} - \tilde{E}^{a[i|}\tilde{E}^{b|j]}A_{ai}A_{bj} + \frac{1}{\gamma}\epsilon_{ijk}A_{a}{}^{i}\tilde{E}^{aj}\tilde{Z}^{k} \Bigg] \notag \\
	& &\hspace{-5mm} - \frac{1}{4} \left(\tilde{f}^{i}{}_{i}\right)^{2} + \frac{1}{2} \tilde{f}^{ij}\tilde{f}_{(ij)} - \frac{1}{2} \tilde{f}_{ij}l^{i}l^{j} \left(\tilde{f}^{k}{}_{k}-\frac{1}{2}\tilde{f}_{kl}l^{k}l^{l}\right) \notag \\
	& &\hspace{-5mm} + \frac{1}{2}\tilde{f}_{(ik)}\left(\tilde{f}^{k}{}_{j}+ \tilde{f}_{j}{}^{k}\right) l^{i} l^{j} \Bigg\rbrace + \left|\left(1+ l_{i}l^{i}\right)\tilde{\tilde{E}}\right| \Lambda,
	\end{eqnarray}
	where $\tilde{\tilde{E}}:=\det(\tilde{E}^{ai})$; we have assumed $\tilde{\tilde{q}}<0$ on the timelike leaves, and all the terms proportional to the Gauss constraints have been neglected to simplify our expressions. We immediately observe that the splitting of the Lorentz indices makes the ensuing formulation look much more complicated than the one obtained by the manifestly Lorentz-covariant approach of Sec.~\ref{sec:FCH}. 
	
	It is worth mentioning that the Barbero-like formulation of Sec.~\ref{sec:GF} can also be obtained from the present description by imposing the gauge condition $l^{i}=0$; solving it together with $\tilde{\mathcal{G}}^{i3}$ yields 
	\begin{equation}
	\tilde{Z}^{i} = - \gamma \partial_{a}{\tilde{E}^{ai}} = -\gamma \epsilon^{ijk}\Gamma_{aj} \tilde{E}^{a}{}_{k},
	\end{equation}
	where, for the last equality to hold, $\Gamma_{ai}$ is the spin connection compatible with $\tilde{E}^{ai}$. Continuing along this path is not as straightforward as neither of the cases presented earlier in Secs. \ref{sec:GF} and \ref{sec:AA}, but we actually obtain the same Barbero-like formulation regardless of the followed approach (modulo the Gauss constraint in the latter, since we have dropped several terms before the gauge fixing).

	\bibliographystyle{apsrev4-1}
	\bibliography{references}

\end{document}